\documentclass{ws-ijmpa}
\usepackage[numbers,sort&compress]{natbib}
\usepackage[english]{babel}
\begin{document}
\newcommand{\DAF}{DA$\Phi$NE}
\newcommand{\Vus}{$V_{us}$}
\newcommand{\Reeps}{\ensuremath{\mathrm{Re}\,(\epsilon'/\epsilon)}}
\newcommand{\Imeps}{\ensuremath{\mathrm{Im}\,(\epsilon'/\epsilon)}}
\newcommand{\gfkk}{$g_{fkk}$}
\newcommand{\gakk}{$g_{akk}$}
\markboth{Authors' Names}
{Instructions for Typing Manuscripts (Paper's Title)}
%
\catchline{}{}{}{}{}
%
\title{PROSPECTS FOR KLOE-2} 
\author{P.~MOSKAL\footnote{e-mail address:
p.moskal@uj.edu.pl}~$^{1}$ FOR AND ON BEHALF OF THE KLOE-2 COLLABORATION\footnote{
F.~Archilli,
D.~Babusci,
D.~Badoni,
G.~Bencivenni,
C.~Bini,
C.~Bloise,
V.~Bocci,
F.~Bossi,
\mbox{P.~Branchini,}
A.~Budano,
S.~A.~Bulychjev,
P.~Campana,
G.~Capon,
\mbox{F.~Ceradini,}
P.~Ciambrone,
E.~Czerwi\'nski,
E.~Dan\'e,
E.~De~Lucia,
\mbox{G.~De~Robertis,}
A.~De~Santis,
G.~De~Zorzi,
A.~Di~Domenico,
\mbox{C.~Di~Donato,}
B.~Di~Micco,
D.~Domenici,
O.~Erriquez,
G.~Felici,
S.~Fiore,
P.~Franzini,
P.~Gauzzi,
S.~Giovannella,
F.~Gonnella,
E.~Graziani,
F.~Happacher,
B.~H\"oistad,
E.~Iarocci,
M.~Jacewicz,
T.~Johansson,
V.~Kulikov,
A.~Kupsc,
J.~Lee-Franzini,
F.~Loddo,
\mbox{M.~Martemianov,}
M.~Martini,
M.~Matsyuk,
R.~Messi,
S.~Miscetti,
\mbox{D.~Moricciani,}
G.~Morello,
P.~Moskal,
F.~Nguyen,
A.~Passeri,
\mbox{V.~Patera,}
A.~Ranieri,
P.~Santangelo,
I.~Sarra,
M.~Schioppa,
\mbox{B.~Sciascia,}
\mbox{A.~Sciubba,}
M.~Silarski,
S.~Stucci,
C.~Taccini,
L.~Tortora,
\mbox{G.~Venanzoni,}
\mbox{R.~Versaci,}
W.~Wi\'slicki,
M.~Wolke,
J.~Zdebik
}
}
\address{
$^1$~Institute of Physics, Jagiellonian University, PL-30-059 Cracow, Poland\\
}
\maketitle
\begin{history}
\received{Day Month Year}
\revised{Day Month Year}
\end{history}
\begin{abstract}
The basic motivation  of the KLOE-2 experiment is
the test of fundamental symmetries and Quantum Mechanics coherence
of the neutral kaon system, and the search for phenomena beyond
the Standard Model in the hadronic and leptonic decays  of ground-state mesons.
Perspectives
for  experimentation by means of  the KLOE-2 apparatus equipped 
with the inner tracker,
new scintillation calorimeters,
and the $\gamma\gamma$ taggers
at the DA$\Phi$NE electron-positron collider upgraded
in luminosity and energy
are presented.

\keywords{
electron-positron annihilation; $\phi$-factory;  kaon interferometry; discrete symmetries; $\gamma\gamma$ physics; scalar spectroscopy; hadronic cross sections; g-2;  $\alpha_{em}$}
\end{abstract}
\ccode{PACS numbers: 11.30.Er, 12.15.Ji, 13.66.Bc, 13.66.Jn, 14.40.Be, 14.40.Df}

\section{introduction}
The KLOE-2 experimental setup~\cite{kloe2,kloe2_fabio} 
is a successor of KLOE~\cite{kloe,kloe_caterina,EMCkloe,DCkloe},
which is at present being upgraded by  
new components
in order to improve its tracking and clustering
capabilities as well as in order
to tag $\gamma\gamma$ fusion processes.
The basic motivation
of the KLOE-2 experiment
is
the test of fundamental symmetries and Quantum Mechanics coherence
of the neutral kaon system, and the search for phenomena beyond
the Standard Model~\cite{kloe2,antonio,fabio,caterina}.
Thanks to the
luminosity upgrade of  DA$\Phi$NE~\cite{zubov,milardi2,raimondi}, 
as well as the installation of new detectors,
KLOE-2 will be able to improve the accuracy  of the
measurement of the K$_S$ mesons and to study the time evolution
of the entangled pairs of neutral kaons with an unprecedented precision.
It is worth mentioning that at present KLOE-2 is a unique facility for
testing  novel ideas on the kaonic quantum eraser~\cite{eraser,eraser2}.
KLOE-2 aims at the
significant improvement of the sensitivity of the tests of the
discrete symmetries
in the decays of K, $\eta$ and $\eta^{\prime}$
mesons beyond the presently achieved limits. In some cases like e.g. the tests of $P$, $C$, or $CP$ symmetries
an improvement by more than one order of magnitude is expected
with an integrated luminosity of 20~fb$^{-1}$
to be achieved within 3-4 years of data taking.
Among other issues the KLOE-2 physics program will include investigations of
(i)  universality of the weak interaction of leptons and quarks,
(ii) lepton universality, 
(iii) the structure of the scalar mesons,
(iv) the meson production via $\gamma \gamma$  interaction,
(v) the muon anomalous magnetic moment, 
(vi) the evolution of the  fine structure constant,  
(vii) the narrow di-lepton resonances in context of the hidden dark matter sector.

Hereafter
the upgrade of the detector system will be briefly presented and for
the comprehensive discussion of all of the above mentioned physics issues 
the interested reader is referred to the recent publication describing in details the physics
program of KLOE-2 at the upgraded DA$\Phi$NE collider~\cite{kloe2}.

\section{Upgrade of instrumentation}
\begin{figure}[h]
\parbox{0.44\textwidth}{\psfig{file=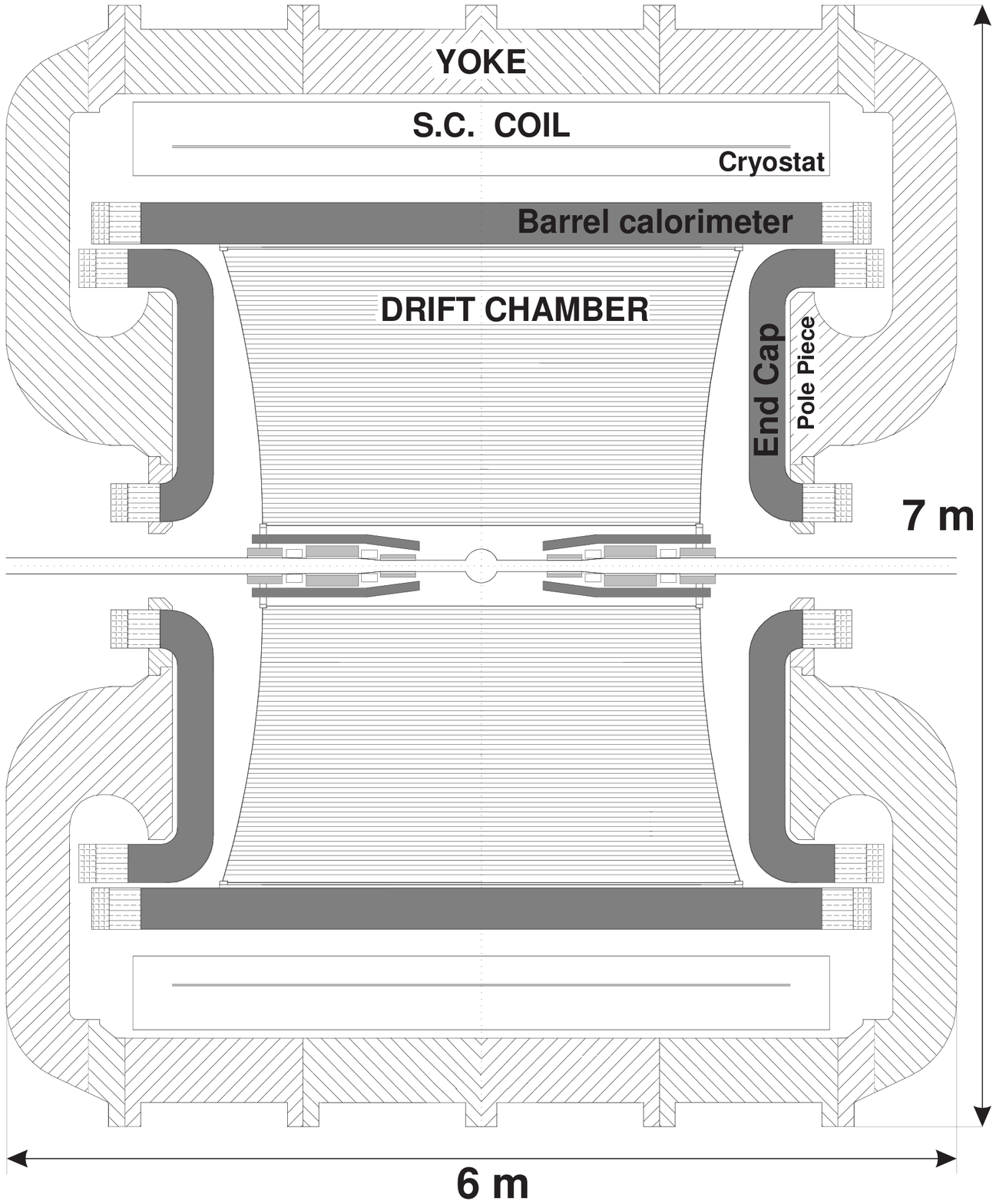,width=5.5cm}}\hspace{-0cm}
\parbox{0.49\textwidth}{\psfig{file=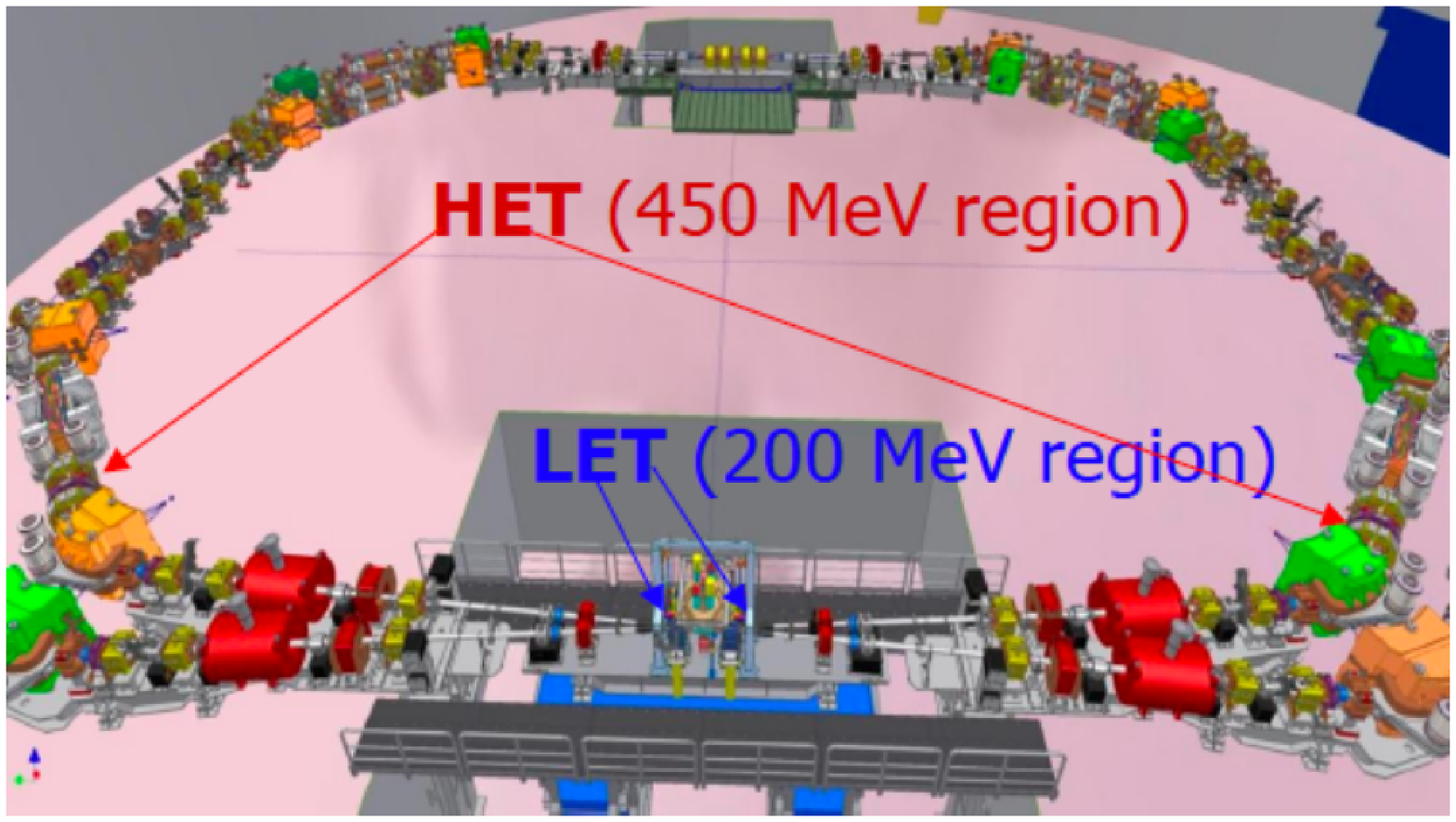,width=7.0cm}}
\caption{ (Left) Cross section of the KLOE detecor.
  (Right) Scheme of the DA${\Phi}$NE collider with positions 
          of Low Energy Taggers (LET) and High Energy Taggers (HET)
          indicated by arrows. The average energy of electrons and positrons covered is also shown. 
  \label{detectors}}
\end{figure}

The Accelerator Division of the INFN Frascati Laboratory
has successfuly commissioned 
the new  electron-positron interaction region, based on 
large Piwinski angle, small beam sizes at the crossing point,  
and {\em Crabbed Waist} compensation of the beam-beam interaction~\cite{zubov,milardi2,raimondi}.
This new solution allowed to increase  the collider
luminosity by a factor of three with respect to the performance reached before 
the upgrade,  and DA$\Phi$NE will deliver up to 15 pb$^{-1}$ per day giving 
possibility to achieve about 20~fb$^{-1}$ within the next 3-4 years of data taking by means of the 
KLOE-2 detector. 
The detector, shown schematically in the left panel of Fig.~\ref{detectors}
consits of a $\sim$~3.5~m long cylindrical drift chamber with a diameter of about 4~m
surrounded by the sampling electromagnetic calorimeter~\cite{kloe,EMCkloe,DCkloe}. 
Both these detectors
are immersed in the axial magnetic field
provided by the superconducting solenoid.  Their functioning and properties have been widely
described 
in previous publications e.g.~\cite{kloe_caterina,kloe,EMCkloe,DCkloe}. Therefore,  hereafter we will only briefly 
discuss the main new components 
of the KLOE-2 setup.

Exclusive measurements of the $\gamma\gamma$ reactions will be enabled by the low
and high energy taggers~\cite{LET} allowing for registration of electrons and positrons 
originating from  $e^{+}e^{-}\to e^{+}e^{-}\gamma^*\gamma^*\to e^{+}e^{-}X$ reaction. 
Right panel of Fig.~\ref{detectors}
shows the scheme of the DA${\Phi}$NE rings with position of taggers indicated by arrows.  

First commissioning runs with KLOE-2 will start next month and,
after collection of statistics corresponding to the integrated luminosity of about 5~fb$^{-1}$,
the next phase of installation of new detectors shall commence 
by the end of the year 2011. 
This stage will include
(i) installation  
of inner tracker~\cite{GEM2010,GEM2010b,GEM2007}
 in order to improve the kaons decay vertex reconstruction and to 
increase the geometrical acceptance for registration of low momentum charged particles 
and (ii) installation of the scintillation calorimeters~\cite{QCALT,QCALT2,CCAL}
in order to increase 
acceptance for registration of very forward electrons and photons outgoing 
from the interaction region
as well as  photons originating from the $K_L$ decays inside the drift chamber.
\begin{figure}[h]
\centerline{\psfig{file=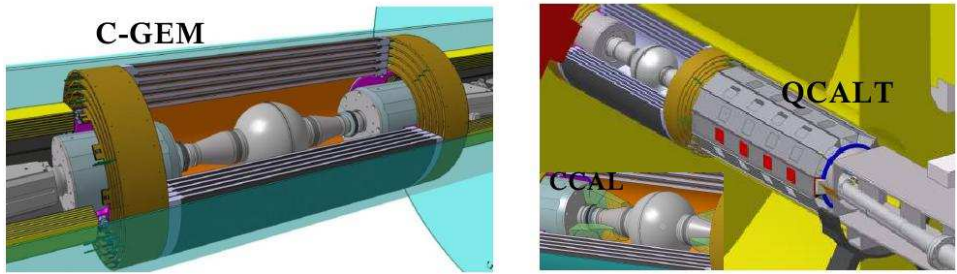,width=\textwidth}}
\caption{Illustration of new detectors which will be installed for the second stage of the KLOE-2 experiment.
  (Left) Inner tracker consisting of four cylindrical layers build out of four gas electron multipliers 
  foils (C-GEM). 
  (Right) Quadrupole Tile Calorimeters (QCALT) and Crystal Calorimeters (CCAL). 
  \label{CGEM_CALS}}
\end{figure}
The new detecors will be placed between beam pipe and the drift chamber as it is indicated in Fig~\ref{CGEM_CALS}.  The CCAL calorimeters will be mounted between the spherical beam-pipe and the quadrupole closest 
to the interaction region whereas the QCALT detectors will sorround the inner quadrupoles. These extra calorimeters will improve the capabilities of the KLOE-2 apparatus e.g. for the search of the 
rare neutral decays of kaons.

In addition to the above mentioned detector upgrades, as described 
in the recent proposal~\cite{upgradeE}, 
we intend also to extend the research towards higher 
electron-positron energies possibly up to 2.5~GeV. An appropriate upgrade of the DA$\Phi$NE 
collider in energy
would allow a precise scan of the multihadronic
cross sections in the energy region where these cross sections are poorly known, 
and hence it would improve significantly 
the accuracy of tests of the Standard Model through a precise
determination of the anomalous magnetic moment of the muon and the
effective fine-structure constant at the $M_Z$ scale. It would enable also 
tests of QCD and effective theories
by investigations of the production 
of all ground state mesons and search for exotics states
in the $\gamma\gamma$ interaction.

\section{Acknowledgements}
The author is grateful to the KLOE-2 Colleagues for the kind help in the preparation 
of the talk for the MESON2010 workshop,
and appreciates corrections of the manuscript by Caterina Bloise, Fabio Bossi  and Antonio Di Domenico.
The author acknowledges also support by INFN and Polish Ministery of Science and 
Higher Education through the Grant No. 0469/B/H03/2009/37.

\end{document}